\documentclass[12pt,draftclsnofoot,onecolumn]{IEEEtran}
\IEEEoverridecommandlockouts
\usepackage{cite,url}
\usepackage{amsmath,amssymb,amsfonts}
\usepackage{algorithmic}
\usepackage{graphicx}
\usepackage{textcomp}
\usepackage{xcolor}
\def\BibTeX{{\rm B\kern-.05em{\sc i\kern-.025em b}\kern-.08em
    T\kern-.1667em\lower.7ex\hbox{E}\kern-.125emX}}

\usepackage{dsfont} %
\usepackage{color}
\usepackage{epstopdf}
\usepackage{hyperref}
\newcommand{\EE}{\mathbb{E}} 

\newcommand{\nv}{{\bf n}}

\newcommand{\xv}{{\bf x}}
\newcommand{\yv}{{\bf y}}

\newcommand{\Hm}{{\bf H}}
\newcommand{\Id}{{\bf I}}

\newcommand{\Mm}{{\bf M}}

\newcommand{\Pm}{{\bf P}}

\newcommand{\Um}{{\bf U}}

\newcommand{\Ec}{{\cal E}}

\newcommand{\Nc}{{\cal N}}

\newcommand{\Uc}{{\cal U}}

\newcommand{\Lambdam}{\boldsymbol{\Lambda}}

\newcommand{\Sigmam}{\boldsymbol{\Sigma}}

\renewcommand{\det}{{\hbox{\rm det}}}

\def\ben{\begin{enumerate}}
\def\beq{\begin{equation}}
\def\beqa{\begin{eqnarray}}
\def\bit{\begin{itemize}}
\def\een{\end{enumerate}}
\def\eeq{\end{equation}}
\def\eeqa{\end{eqnarray}}
\def\eit{\end{itemize}}

\newtheorem{theorem}{Theorem}[section]

\newtheorem{proposition}[theorem]{Proposition}

\DeclareMathOperator{\tr}{tr}

\newcommand{{\diag}}{\rm{diag}}

\newcommand{\eins}{\leavevmode\hbox{\small1\kern-3.8pt\normalsize1}}
\begin{document}

\title{Closed-form performance analysis of linear MIMO receivers in general fading scenarios
\thanks{G. Akemann and M. Kieburg acknowledge support by the
German research council (DFG) through CRC 1283:
“Taming uncertainty and profiting from randomness and
low regularity in analysis, stochastics and their applications". G. Alfano wants to thank the Bielefeld University for its hospitality.}
}

\author{\IEEEauthorblockN{Mario Kieburg \& Gernot Akemann}\\
\IEEEauthorblockA{\textit{Fakultet f\"{u}r Physik} \\
\textit{Universit\"{a}t Bielefeld}\\
Bielefeld, Germany \\
\{mkieburg,akemann\}@physik.uni-bielefeld.de}\\
\and
\IEEEauthorblockN{Giusi Alfano \& Giuseppe Caire}\\
\IEEEauthorblockA{\textit{CommIT Group} \\
\textit{Technische Universit\"{a}t Berlin}\\
Berlin, Germany \\
\{alfano,caire\}@tu-berlin.de}
}

\maketitle

\begin{abstract}
Linear precoding and post-processing schemes are ubiquitous in wireless multi-input-multi-output (MIMO) settings, due to their reduced complexity with respect to optimal strategies. Despite their popularity, the performance analysis of linear MIMO receivers is mostly not available in closed form, apart for the canonical (uncorrelated Rayleigh fading) case, while for more general fading conditions only bounds  are provided. This lack of results is motivated by the complex dependence of the output signal-to-interference and noise ratio (SINR) at each branch of the receiving filter on both the squared singular values as well as the (typically right) singular vectors of the channel matrix. While the explicit knowledge of the statistics of the SINR can be circumvented for some fading types in the analysis of the linear Minimum Mean-Squared Error (MMSE) receiver, this does not apply to the less complex and widely adopted Zero-Forcing (ZF) scheme. This work provides the first-to-date closed-form expression of the probability density function (pdf) of the output ZF and MMSE SINR, for a wide range of fading laws, encompassing,  in particular, correlations and multiple scattering effects typical of practically relevant channel models.
\end{abstract}

\begin{IEEEkeywords}
Linear Receivers; ZF; MMSE; MIMO; Multiple Scattering; Meijer G Function
\end{IEEEkeywords}

\section{Introduction}

Sum rate analysis of multiantenna systems operating in fading scenarios with  linear  processing at the receiver plays an important role in the design of modern MIMO wireless communication systems, due to the fact that these receivers offer near-optimal performance in  many relevant scenarios at much lower complexity than their maximum-likelihood counterpart.
For a finite number of antennas, the performance of linear Zero-Forcing (ZF) and Minimum Mean-Square Error (MMSE) receivers has been characterized in closed form in \cite{ZFGamma} and in \cite{Gaommse}, respectively, and then compared in the asymptotic regime of high input Signal-to-Noise-Ratio (SNR) in \cite{JiangVara}.
Such closed-form analysis is restricted to the simple i.i.d. Rayleigh fading statistics, where the elements of the channel vectors are i.i.d. complex Gaussian circularly symmetric coefficients with the same variance. For more complicated (and yet more relevant in practical applications) statistics,
only semi-analytic bounds are available, requiring the evaluation of certain quantities via Monte Carlo simulation. As an alternative, the large-system asymptotic limit has been widely investigated for a great variety of fading statistics, and closed-form or quasi-closed-form expressions for the achievable spectral efficiency have been characterized in the limit of number of receive and transmit antennas going to infinity at fixed ratio (see e.g. \cite[and references therein]{lin_Caire09} for a detailed list of results in the large-system settings).

 While ZF performance analysis has been extended to Ricean fading in \cite{SiriteZF,SiriteSchur}, only a (multi-dimensional) integral expression is to date available for the Rayleigh (also with one-sided correlation) SINR probability density function in the  MMSE case \cite{ParkI,ParkII}; such  expressions lead however to explicit formulae only for very low number of antennas, below the currently exploited maximum per-device antenna number.  ZF receivers and precoders are otherwise widely adopted in the massive MIMO setting, but classical expressions of the SINR, relating its value to the norm of the columns of the channel matrix, rather than directly  to the spectral properties of such matrix, usually yield to cumbersome expressions \cite{PNR17}, tailored to a specific fading law \cite{TSGD17}.

In absence of explicit SINR characterization for most of the fading models of interest, the sum rate analysis has been mainly obtained by direct study of the rate expression. In \cite{Mckaymmse}, a closed-form expression of the sum rate achievable by an MMSE receiver on a MIMO link on a fading channel whose matrix has independent columns is provided, exploiting elementary properties of the determinant and linking the sum rate to the ergodic mutual information.
Only bounds, tailored to specific transmit SNR values, have been provided instead for the ZF case, see e.g. \cite{MichalisZF,Michalisglob}, for the most common fading models (Rayleigh with or without correlation, Rice). In \cite{2018TIT}, considering a  multiple-cluster scattering channel,  a closed-form expression for the MMSE sum rate has been derived, relying on the findings of \cite{Mckaymmse}; however, in this case, ZF-related bounds, elsewhere showed to be very tight by means of numerical simulation (see again \cite{MichalisZF,Michalisglob}), turn out to be quite loose in the multiple scattering case.

In this paper we (partially) fill the gap in the finite-dimensional analysis of linear MIMO receiver schemes for a variety of relevant fading statistics. In particular, we investigate the statistics of the output SINR  and analytically evaluate the sum rate for a broad class of fading laws and for arbitrary number of antennas. The study is conducted by borrowing mathematical tools recently developed in random matrix theory by two of the co-authors \cite{AIK2013,KK2016}, and leads to closed-form expressions, numerically computable with built-in functions by widely adopted symbolic calculus software tools.

The work is articulated as follows: section~\ref{sec:model} introduces the considered system and channel matrix model. In section~\ref{sec:analytics} we provide the statistical characterization of the sum rate and the distribution of the SINR for  the LMMSE and the ZF receiver.
A numerical example is discussed in Section~\ref{sec:numerics}, and in section~\ref{sec:conclusio} we summarize and give an outlook where our results may be applied to.
{\em Notation:} Vectors are denoted by boldface lowercase letters, matrices by boldface uppercase; $\partial_x$ is used as a shortcut for $\frac{\partial}{\partial x}$.

\section{System and Channel Model}\label{sec:model}

A multi-input multi-output (MIMO)
system with $n_t$ transmitters and $n_r\ge n_t$ receivers can be  represented by the
input-output relationship:
 \begin{equation}
\label{eq:model}
\yv = \sqrt{\alpha}\Hm\xv + \nv\,,
\end{equation}
where the received signal vector $\yv$ is of length $n_r$, $\Hm$ is the
$n_r\times n_t$ random channel matrix, $\xv$ is a random input vector of size
$n_t$ with covariance $\EE[\xv\xv^\dagger]=\Ec_s/n_t\Id$, and $\nv$ represents
Gaussian noise with covariance
$\EE[\nv\nv^\dagger]=\Nc_0\Id$. As an energy-normalization constraint we define the factor $\alpha=\frac{n_t\,n_r}{\EE\{\tr\{\Hm^\dagger\Hm\}\}}$.

In case of independent stream decoding, the
output SINR corresponding to the $k$-th transmitted signal stream can be
expressed for the MMSE and, respectively, for the ZF receiver as\footnote{In \cite{Verdu}, the acronym SIR is used, rather than SINR.}
\cite[Ch. 6]{Verdu}:
\begin{equation}\label{ithmode}
\gamma^{\rm mmse}_{k}\mathord{=}\frac{1}{\left[\left(\Id\mathord{+}\delta\Hm^\dagger\Hm\right)^{-1}\right]_{k,k}}\mathord{-}1,\
\gamma^{\rm zf}_{k}\mathord{=}\frac{\delta}{\left[\left(\Hm^\dagger\Hm\right)^{-1}\right]_{k,k}}\,,
\end{equation}
where $\delta=\frac{\Ec_s\alpha}{n_t}$ takes into account both the transmit power level as well as the power-normalization constraint depending on the fading law. The expression of the ergodic sum rate, in the case of perfect channel state information (CSI) knowledge at the receiver and statistical knowledge at the transmitter, achievable by a linear receiver on a MIMO channel can be derived by its distribution as the expectation
\begin{equation}\label{def:sum-rate}
R\triangleq \sum_{k=1}^{n_t}\EE[\ln(1+\gamma_k)]\,,
\end{equation}
where $\gamma_k$ is to be particularized to the receiver in use.
Its statistics depend on a single diagonal element of the inverse of a Hermitian random matrix that is given in terms of the channel matrix $\Hm$ as in~\eqref{ithmode}.

Most of the previous works approached the problem of characterizing the law of \eqref{ithmode} resorting to an expression of the SINR in terms of a quadratic form with random kernel and vector given by the corresponding column of the channel matrix. Such an approach, while providing insights in the asymptotic case, fails to offer a closed-form expression for the statistics of $\gamma_k$ whenever the columns of the channel matrix do not follow a complex Gaussian law with zero mean. Our technique, based on the spectral properties of the channel matrix, provides instead an exact result for a class of random matrices encompassing most of the  fading laws for MIMO systems adopted in the literature.

\section{SINR and Sum Rate Characterization for Polynomial Ensembles}\label{sec:analytics}

The closed expressions of the density of $\gamma_k$ obtained in this work hold for the class of random matrices usually referred to as~\emph{invariant polynomial ensembles} (IPE)~\cite{KS2014,KK2016}. Recalling that we assume $n_r\ge n_t$, an IPE is best described in terms of the eigenvalue decomposition $\Hm^\dagger\Hm=\Um^\dagger\Lambdam\Um$. The unitary matrix $\Um\in\mathcal{U}(n_t)$ is  Haar-distributed and the joint law of the eigenvalues $\Lambdam=\diag(\lambda_1,\ldots,\lambda_{n_t})>0$ has the form
\begin{equation}\label{jpdf:polyn}
p(\Lambdam)=\frac{1}{{n_t}!}\Delta_{n_t}(\Lambdam)\det[q_{b-1}(\lambda_a)]_{a,b=1,\ldots,n_t}.
\end{equation}
The eigenvalues are  distinct w.p.1, $\Delta_{n_t}(\Lambdam)=\prod_{k>\ell=1}^{n_t}(\lambda_k-\lambda_\ell)$ is the Vandermonde determinant in the entries of $\Lambdam$,  and   $q_j$ are some deterministic biorthonormal functions. 

Many common random matrix models are included in the class of IPEs, e.g., the centered (uncorrelated) and the one-sided correlated (at the receiver) Wishart ensemble~\cite{win}, and products of normal distributed rectangular matrices~\cite{Muller}, see also~\cite{Palomar} for a detailed summary of matrix models. For most of these examples we give explicit expressions for the distribution of the output SINR and the corresponding mean sum rate.
Our strategy is to make use of the biorthogonal structure associated to the joint law~\eqref{jpdf:polyn}. Due to the determinantal expression of \eqref{jpdf:polyn},  we can identify monic polynomials $p_j$ of order $j$, that  meet the following orthonormality constraint w.r.t. the $q_j$ functions appearing in the joint eigenvalues density above:
\begin{equation}\label{biorth}
\int_0^\infty p_a(\lambda)q_b(\lambda)d\lambda=\delta_{ab},\quad a,b=0,\ldots,n_t-1.
\end{equation}
By virtue of the orthogonality constraint above, \eqref{jpdf:polyn} can be equivalently written as
\begin{equation}\label{jpdf:polyn2}
p(\Lambdam)=\frac{1}{n_t!}\det[p_{b-1}(\lambda_a)]_{a,b=1,\ldots,n_t}\det[q_{b-1}(\lambda_a)]_{a,b=1,\ldots,n_t}\,.
\end{equation}
This results in the following expression for the joint density of any unordered subset of $k\leq\,n_t$ eigenvalues (otherwise referred to as $k$-point correlation function in physics)
\begin{equation}\label{kpoint}
R_k(\lambda_1,\ldots,\lambda_k)=\frac{n_t!}{(n_t-k)!}\int d\lambda_{k+1}\cdots d\lambda_{n_t}p(\Lambdam)
=\det[K(\lambda_a,\lambda_b)]_{a,b=1,\ldots,n_t}
\end{equation}
 depending on the kernel
\begin{equation}\label{kernel}
K(\lambda_a,\lambda_b)={\sum}_{j=0}^{n_t-1}p_j(\lambda_a)q_j(\lambda_b).
\end{equation}
This analytical structure carries over to the joint law of the eigenvalues of $(\Hm^\dagger\Hm)^{-1}$ and $\left(\Id\mathord{+}\delta\Hm^\dagger\Hm\right)^{-1}$, for IPE matrices\footnote{Later in our work, some extensions beyond IPE will be also discussed.}.

\subsection{ZF Receiver}

Let us focus on  ZF receivers first.  Throughout the work, we shall make an extensive use of  Meijer G-functions that are defined as contour integrals~\cite{Gradshteyn}
\begin{equation}\label{MeijerG}
\begin{split}
&G_{n+p,m+q}^{m,n}\left(\left.\begin{array}{c} a_1,\ldots,a_n;b_1,\ldots,b_p \\ c_1,\ldots,c_m;d_1,\ldots,d_q\end{array}\right|x\right)\\
=&\int_{\mathcal{C}}\frac{ds}{2\pi i} \frac{\left(\prod_{j=1}^m \Gamma(c_j-s)\right)\left(\prod_{j=1}^n \Gamma(1-a_j+s)\right)}{\left(\prod_{j=1}^p\Gamma(b_j-s)\right)\left(\prod_{j=1}^q\Gamma(1-d_j+s)\right)}x^{s},
\end{split}
\end{equation}
where the contour runs from $-i\infty$ to $+i\infty$ and has the poles of $\Gamma(c_j-s)$ on the right hand side of the path and of $\Gamma(1-a_j+s)$ on the left hand side.

\begin{proposition}\label{prop:ZF}
Let $n_t\geq2$ and the channel matrix $\Hm=\Pm\sqrt{\Sigmam_t}$,  be an IPE with the kernel~\eqref{kernel}, $\Pm$ a random matrix  and $\Sigmam_t>0$ representing the correlation matrix  at the transmitter. Moreover define $\sigma_k=(\Sigmam_t^{-1})_{kk}$.
The distribution of the output SINR of the $k$th transmitter is given by
\begin{equation}\label{distributionZF}
\rho(\gamma^{\rm zf}_{k})=(n_t-1)\frac{\sigma_k}{\delta}\int_{0}^1 dx K\left(0,\frac{\sigma_k\gamma^{\rm zf}_kx}{\delta}\right)\left(1-x\right)^{n_t-2}x
\end{equation}
for a ZF receiver and the corresponding sum rate is
\begin{equation}\label{sumrateZF}
R=(n_t-1)!\int_0^\infty dx K(0,x)\sum_{k=1}^{n_t}G_{3,3}^{2,2}\left(\left.\begin{array}{c} 1,1;n_t \\ 1,1;\,0\end{array}\right|\frac{\delta }{\sigma_k} x\right).
\end{equation}
\end{proposition}

\begin{IEEEproof}
We start with the fact that $[(\Hm^\dagger\Hm)^{-1}]_{k,k}=v_k^\dagger(\Pm^\dagger\Pm)^{-1}v_k$ where $v_k=\Sigmam^{-1/2}e_k$ and $e_k$ is the $n_t$-dimensional unit vector with unity at the $k$th position and elsewhere zero. Since the eigenvectors of $(\Pm^\dagger\Pm)^{-1}$ are drawn from the Haar measure $d\mu(\Um)$ of $\mathcal{U}(n_t)$ we know that $[(\Hm^\dagger\Hm)^{-1}]_{k,k}$ and $\sigma_k[(\Pm^\dagger\Pm)^{-1}]_{1,1}$ share the same distribution. Hence the characteristic function of $(\gamma_k^{\rm zf})^{-1}$ is
\begin{equation}\label{Fourier.a}
\begin{split}
\mathcal{L}(s)=\int_{\mathbb{R}_+^n} d\Lambdam\int_{{\Uc}(n)}d\mu(\Um)p(\Lambdam)\exp\left[-\frac{s\sigma_k}{\delta}e_1^T\Um^\dagger \Lambdam^{-1} \Um e_1\right],
\end{split}
\end{equation}
where we already decomposed $\Pm^\dagger\Pm=\Um^\dagger\Lambdam\Um$ with $\Lambdam>0$ distributed by~\eqref{jpdf:polyn2} and $\Um\in\mathcal{U}(n_t)$. The integral over the unitary group can be done with the Harish-Chandra--Itzykson--Zuber integral~\cite{HC,IZ}
\begin{equation}\label{HCIZ}
\begin{split}
&\int_{{\Uc}(n)}d\mu(U)\exp\left[-\frac{s\sigma_k}{\delta}e_1^T\Um^\dagger \Lambdam^{-1} \Um e_1\right]\\
=&\frac{(n_t-1)!}{(-s\sigma_k/\delta)^{n_t-1}\Delta_{n_t}(\Lambdam^{-1})}\det[ \lambda_a^{1-b},\ e^{-s\sigma_k/(\lambda_a\delta)}]_{\substack{a=1\ldots,n_t\\b=1,\ldots,n_t-1}}.
\end{split}
\end{equation}
The ratio of the two Vandermonde determinants in~\eqref{jpdf:polyn} and~\eqref{HCIZ} yields $\Delta_{n_t}(\Lambdam)/\Delta_{n_t}(\Lambdam^{-1})=(-1)^{n_t(n_t-1)/2}\det\Lambdam^{n_t-1}$. This term can be combined with the second determinant in the numerator of~\eqref{jpdf:polyn} by pulling the determinant into the rows  and reshuffling the columns.
Next we recombine the first $n_t-1$ rows of this determinant into the polynomials $\tilde{p}_{b-1}(\lambda_a)=p_{b-1}(\lambda_a)-p_{b-1}(0)$, see~\eqref{jpdf:polyn2}, where the subtraction of the constant is due to the missing monomial of order zero in these columns.
The kernel $K(0,\lambda)$, see~\eqref{kernel}, is orthogonal to $\tilde{p}_{b-1}(\lambda_a)$ for all $b=2,\ldots,n_t$. Thus, we choose the new weight $K(0,\lambda)$  instead
 of $q_0(\lambda)$,  by taking linear combinations in the second determinant of \eqref{jpdf:polyn}.
Applying a generalization of Andreief's identity~\cite{and83,KG2010},  we obtain the
 determinant of a diagonal matrix,  that leads to
\begin{equation}\label{Fourier.b}
\begin{split}
\mathcal{L}(s)=(n_t-1)!\int_0^\infty dx K(0,x)\left(-\frac{x\delta}{s\sigma_k}\right)^{n_t-1}e^{-\frac{s\sigma_k}{x\delta}}.
\end{split}
\end{equation}
The inverse Laplace transform yields the distribution of $(\gamma_k^{\rm zf})^{-1}$ and can be performed for the $s$ dependent part in the integrand with the help of the residue theorem. Hence, the distribution of $\gamma_k^{\rm zf}$ is
\begin{equation}\label{dens.inv}
\begin{split}
\rho(\gamma_k^{\rm zf})=\frac{(1-n_t)}{(\gamma_k^{\rm zf})^2}\int_{\sigma_k\gamma_k/\delta}^\infty dx K(0,x)\left(1-\frac{x\delta}{\gamma_k^{\rm zf}\sigma_k}\right)^{n_t-2}\frac{x\delta}{\sigma_k}.
\end{split}
\end{equation}
When employing the fact that $K(0,x)$ is orthogonal to all monomials from order one to $n_t-1$, we can switch to an integral on the interval $[0,\sigma_k\gamma_k/\delta]$ instead of $[\sigma_k\gamma_k/\delta,\infty[$. Additionally we rescale $x\to\gamma_k\sigma_k/\delta$ and we arrive at~\eqref{distributionZF}.

The sum rate can be obtained by the following identities of two particular Meijer G-functions
\begin{align}
 {\rm ln}\left(1+x\right)=&G_{2,2}^{1,2}\left(\left.\begin{array}{c} 1,1;\, - \\ 1;0\end{array}\right|x\right),\label{log}\\
(1-x)^{n_t-2}x\Theta(1-x)=&(n_t-2)!G_{1,1}^{1,0}\left(\left.\begin{array}{c} -\,;n_t \\ 1;\,-\end{array}\right|x\right)
\end{align}
with $\Theta(x)$ denoting the  step function. Then, the Mellin convolution of two Meijer G-functions, see~\cite[(7.811.1)]{Gradshteyn}, leads to our second claim~\eqref{sumrateZF}.
\end{IEEEproof}

In order to illustrate this result, let us consider some more explicit special cases. In the first example, we  consider the situation where $\Pm=\Pm_M\cdots\Pm_1$ is a product of $M$ normal distributed complex random matrices, as considered in~\cite{AKW2013,WZCT,2018TIT}. The dimension of the matrix $\Pm_j$ is assumed to be $(n_t+\nu_{j})\times(n_t+\nu_{j-1})$ with $\nu_0=0$ and $\nu_j\geq0$ for $j=1,\ldots,M$. The product matrix $\Pm$ belongs to the class of P\'olya Ensembles~\cite{KK2016,FKK2017} (PE) whose statistics is completely determined by a single function $\omega(x)$ as follows~\cite{KK2016}
\begin{align}
p_l(\lambda)=&\sum_{j=0}^l\frac{(-1)^{l-j}l!\mathcal{M}\omega(l+1)}{j!(l-j)!\mathcal{M}\omega(j+1)}\lambda^j, \label{weightp}\\
q_{l}(\lambda)=&\frac{1}{l! \mathcal{M}\omega(l+1)}\partial_{\lambda}^l\left[(-\lambda)^l\omega(\lambda)\right],\label{weightq}\\
K(x,y)=&-n_t\frac{\mathcal{M}\omega(n_t+1)}{\mathcal{M}\omega(n_t)}\int_0^1p_{n_t-1}(xu)q_{n_t}(yu)du \label{kernelPolya}
\end{align}
with $\mathcal{M}\omega(s)$ the Mellin transform of $\omega(x)$. As a result of this particular structure one can readily derive
\begin{equation}\label{Poly.kern}
K(0,y)=\frac{1}{(n_t-1)!\mathcal{M}\omega(1)}\frac{1}{y}\partial_y^{n_t-1}[y^{n_t}\omega(y)]
\end{equation}
and, thus, for the distribution of the SINR
\begin{equation}\label{PolyadistributionZF}
\rho_{\rm P\text{\'o}lya}(\gamma^{\rm zf}_{k})=\frac{\sigma_k}{\delta\mathcal{M}\omega(1)}\omega\left(\frac{\sigma_k\gamma^{\rm zf}_k}{\delta}\right)
\end{equation}
and for the sum rate
\begin{equation}\label{PolyasumrateZF}
R_{\rm P\text{\'o}lya}=\frac{1}{\mathcal{M}\omega(1)}\int_0^\infty dx \omega(x)\sum_{k=1}^{n_t}G_{2,2}^{1,2}\left(\left.\begin{array}{c} 1,1;\,- \\ 1;\,0\end{array}\right|\frac{\delta }{\sigma_k} x\right).
\end{equation}
Here, we have applied integration by parts and employed the contour representation~\eqref{MeijerG} when applying the derivative on the Meijer G-function. Interestingly, the distribution of the SINR is, apart from $\delta$, completely independent of the matrix size and the sum rate is exactly linear in the number of antennas $n_t$ when $\Sigmam_t=\Id$, independently of the weight $\omega(x)$.
For a  channel matrix with a non-trivial transmit correlation matrix, i.e. $\Hm=\Pm\sqrt{\Sigmam_t}=\Pm_M\cdots\Pm_1\sqrt{\Sigmam_t}$,
one can write, according to~\cite{AKW2013,AIK2013,KK2016}
\begin{equation}
\begin{split}
\omega_{\rm Gauss}(x)=&G_{0,M}^{M,0}\left(\left.\begin{array}{c} -\,;\, - \\ \nu_1,\ldots,\nu_M;\,-\end{array}\right|x\right).
\end{split}
\end{equation}
 and the Mellin transform of the weight at $x=1$ is
\begin{equation}
\mathcal{M}\omega_{\rm Gauss}(1)=\prod_{j=1}^M\nu_j!.
\end{equation}
Therefore, the corresponding SINR distribution is given by
\begin{equation}\label{ProddistributionZF}
\rho_{M}(\gamma^{\rm zf}_{k})=\frac{\sigma_k}{\left(\prod_{j=1}^M\nu_j!\right)\delta}
G_{0,M}^{M,0}\left(\left.\begin{array}{c} -\,;\, - \\ \nu_1,\ldots,\nu_M;\,-\end{array}\right|\frac{\sigma_k\gamma^{\rm zf}_k}{\delta}\right)
\end{equation}
and the resulting sum rate is
\begin{equation}\label{ProdsumrateZF}
R_{M}=\frac{1}{\prod_{j=1}^M\nu_j!}
\sum_{k=1}^{n_t}G_{M+2,2}^{1,M+2}\left(\left.\begin{array}{c} -\nu_1,\ldots,-\nu_M,1,1;\,- \\ 1;\,0\end{array}\right|\frac{\delta }{\sigma_k} \right).
\end{equation}
where we resort again to the Mellin convolution identity \cite[(7.811.1)]{Gradshteyn}.
The above expressions provide the SINR density and expected sum rate of a MIMO system with multiple-cluster (or progressive) scattering constituted by $M$ independent scattering layers, whose performance in presence of linear receivers have been previously only investigated in~\cite{2018TIT}, in absence of spatial correlation and not in closed form. As already remarked in the introduction, most widely used bounding techniques may perform poorly in the case of  ZF sum rate in presence of a product channel.

For $M=1$ the results simplify further  because $\omega_{\rm Gauss}(x)=x^\nu e^{-x}$ so that we have
\begin{equation}\label{M1distributionZF}
\begin{split}
\rho_{M=1}(\gamma^{\rm zf}_{k})=&\frac{\sigma_k}{\nu!\delta}\left(\frac{\sigma_k\gamma^{\rm zf}_k}{\delta}\right)^\nu\exp\left[-\frac{\sigma_k\gamma^{\rm zf}_k}{\delta}\right]
\end{split}
\end{equation}
and its corresponding sum rate can be simplified to
\begin{equation}\label{M1sumrateZF}
\begin{split}
R_{M=1}=\frac{1}{\nu!}\sum_{k=1}^{n_t}G_{3,2}^{1,3}\left(\left.\begin{array}{c} -\nu,1,1;\,- \\ 1;\,0\end{array}\right|\frac{\delta }{\sigma_k} \right).
\end{split}
\end{equation}

As a second example we consider the PE $\Pm=\Pm_M\cdots \Pm_1$ with $\Pm_j\in\mathbb{C}^{(n+\nu_{j-1})\times(n+\nu_{j})}$ the truncation of a $(n+\mu_j)\times (n+\mu_j)$ unitary matrix with $\mu_j\geq\nu_j,\nu_{j-1}$, which are also known as Jacobi ensemble~\cite{KKS2015,KK2016}. The corresponding weight is~\cite{KKS2015,KK2016}
\begin{equation}
\begin{split}
\omega_{\rm Jacobi}(x)=&G_{M,M}^{M,0}\left(\left.\begin{array}{c} -\,;\mu_1,\ldots,\mu_M \\ \nu_1,\ldots,\nu_M;\,-\end{array}\right|x\right)
\end{split}
\end{equation}
and its Mellin transform at $1$ becomes
\begin{equation}
\mathcal{M}\omega_{\rm Jacobi}(1)=\prod_{j=1}^M\frac{\nu_j!}{\mu_j!}.
\end{equation}
Therefore, we get
\begin{equation}\label{density.prod.Jac}
\rho_{\rm Jacobi}(\gamma^{\rm zf}_{k})=\frac{\sigma_k}{\delta}\left(\prod_{j=1}^M\frac{\mu_j!}{\nu_j!}\right)
G_{M,M}^{M,0}\left(\left.\begin{array}{c} -\,;\mu_1,\ldots,\mu_M \\ \nu_1,\ldots,\nu_M;\,-\end{array}\right|\frac{\sigma_k\gamma^{\rm zf}_k}{\delta}\right)
\end{equation}
and
\begin{equation}\label{res:sumrate.prod.Jac}
R_{\rm Jacobi}=\prod_{j=1}^M\frac{\mu_j!}{\nu_j!}\sum_{k=1}^{n_t} G_{M+2,M+2}^{1,M+2}\left(\left.\begin{array}{c} -\nu_1,\ldots,-\nu_M,1,1;\,- \\ 1;-\mu_1,\ldots,-\mu_M,0\end{array}\right|\frac{\delta }{\sigma_k}\right).
\end{equation}

The Jacobi model has received considerable attention in the last few years, since it adequately models a multi-channel communication in optical fibres  \cite{JacobiMIMO,KMV};  ZF detection is adopted in this framework only in presence of weakly coupled optical fiber modes.

\subsection{MMSE Receiver}

In this section we focus on the performance analysis of the MMSE receiver, for which we can only provide results for the case of no transmit correlation. The treatment of a non-trivial transmit correlation is deferred to future work.

\begin{proposition}\label{prop:MMSE}
Let $n_t\geq2$ and the channel matrix $\Hm$ be a PE with the kernel~\eqref{kernel} with a trivial correlation at transmitter, i.e., $\Sigmam_t=\Id$.
Then, the distribution of the output SINR of the $k$th transmitter is given by
\begin{equation}\label{distributionMMSE}
\rho(\gamma^{\rm mmse})=\frac{n_t-1}{\delta}\left(\frac{\gamma^{\rm mmse}}{\gamma^{\rm mmse}+1}\right)^{n_t}\int_{0}^1 dx\left(1-x\right)^{n_t-2} \left(x+\frac{1}{\gamma^{\rm mmse}}\right)K\left(-\frac{1}{\delta},\frac{\gamma^{\rm mmse}}{\delta}x\right)
\end{equation}
for an MMSE receiver and the associated sum rate is
\begin{equation}\label{sumrateMMSE}
\begin{split}
R=&n_t\biggl[\int_{0}^\infty dx\,{\rm ln}(1+ x\,\delta)K\left(-\frac{1}{\delta},x\right)+\gamma+\Psi(n_t)\biggl]
\end{split}
\end{equation}
with $\gamma$ the Euler constant and $\Psi(x)$ the Digamma function~\cite{Gradshteyn}
\end{proposition}

\begin{IEEEproof}
Since there is no transmit correlation, the SINR is symmetric with respect to the data stream index $k$, and therefore we can simply denote it as $\gamma^{\rm mmse}$, dropping the subscript.

We begin with describing the distribution of $(\gamma^{\rm mmse}+1)^{-1}$ which is given by $[(\Id+\delta\Hm^\dagger\Hm)^{-1}]_{1,1}$ or by any other diagonal element. As before we approach this problem via its  characteristic function. Here, we use the fact that the matrix $\Id+\delta\Hm^\dagger\Hm$ is also an IPE,  with  eigenvalues $\Mm=\diag(\mu_1,\ldots,\mu_{n_t})>0$ related to $\Lambdam$ as  $\mu_j=1+\delta \lambda_j$. Thence, the joint law of $\Mm$ is
\begin{equation}\label{jpdf:polyn.3}
p(\Mm)=\frac{\Delta_{n_t}(\Mm)}{{n_t}!\delta^{n_t(n_t+1)/2}}\det\left[q_{b-1}\left(\frac{\mu_a-1}{\delta}\right)\Theta(\mu_a-1)\right]_{a,b=1,\ldots,n_t}.
\end{equation}
The polynomials bi-orthogonal to the new weight functions $\widetilde{q}_{b-1}(\mu_a)=\delta^{-b}q_{b-1}\left((\mu_a-1)/\delta\right)\Theta(\mu_a-1)$ can be also expressed in terms of the polynomials $p_j$'s, in \eqref{biorth} too, and are $\widetilde{p}_{b-1}(\mu_a)=\delta^{b-1}p_{b-1}\left((\mu_a-1)/\delta\right)$. Therefore, the new kernel has the form
\begin{equation}
\begin{split}
\widetilde{K}(x,y)=\sum_{j=0}^{n_t-1}\widetilde{p}_{j}(x)\widetilde{q}_{j}(y)=\frac{\Theta(y-1)}{\delta}K\left(\frac{x-1}{\delta},\frac{y-1}{\delta}\right).
\end{split}
\end{equation}
Now we can apply Proposition~\ref{prop:ZF}, where we set $\sigma_k=1$ in~\eqref{distributionZF} leading to the distribution of $\gamma^{\rm mmse}$,
\begin{equation}\label{distributionMMSE.b}
\rho(\gamma^{\rm mmse})=\frac{n_t-1}{\delta}\int_{1/(\gamma^{\rm mmse}+1)}^1 dx\left(1-x\right)^{n_t-2}x
K\left(-\frac{1}{\delta},\frac{(\gamma^{\rm mmse}+1)x-1}{\delta}\right)
\end{equation}
When substituting $x\to(\gamma^{\rm mmse} x+1)/(\gamma^{\rm mmse}+1)$ we find our result~\eqref{distributionMMSE}.

For the sum rate, we rescale $x\to x/(1+\gamma^{\rm mmse})$ in~\eqref{distributionMMSE.b}. After interchanging the integrals of $\gamma^{\rm mmse}$ and $x$, we have to perform
\begin{equation}
f=(n_t-1)\int_0^\infty d\gamma^{\rm mmse}\frac{x\,{\rm ln}(1+\gamma^{\rm mmse})}{(1+\gamma^{\rm mmse})^2}
\left(1-\frac{x}{1+\gamma^{\rm mmse}}\right)^{n_t-2}\Theta\left(1-\frac{x}{1+\gamma^{\rm mmse}}\right).
\end{equation}
The latter three terms can be expressed as a derivative in $\gamma^{\rm mmse}$,
\begin{equation}
\begin{split}
&\frac{(n_t-1)x}{(1+\gamma^{\rm mmse})^2}\left(1-\frac{x}{1+\gamma^{\rm mmse}}\right)^{n_t-2}\Theta\left(1-\frac{x}{1+\gamma^{\rm mmse}}\right)\\
=&\partial_{\gamma^{\rm mmse}}\biggl[\left(1-\frac{x}{1+\gamma^{\rm mmse}}\right)^{n_t-1}\Theta\left(1-\frac{x}{1+\gamma^{\rm mmse}}\right)-1\biggl].
\end{split}
\end{equation}
The difference with unity allows an integration by parts since the boundary term at infinity vanishes. Then, we find
\begin{equation}
f=\int_0^\infty \frac{d\gamma^{\rm mmse}}{1+\gamma^{\rm mmse}} \biggl[1
-\left(1-\frac{x}{1+\gamma^{\rm mmse}}\right)^{n_t-1}\Theta\left(1-\frac{x}{1+\gamma^{\rm mmse}}\right)\biggl].
\end{equation}
In the next step we split the integration domain into the intervals $[0,x-1]$ and $[x-1,\infty[$. The integration over the first interval yields ${\rm ln}(x)$ while for the latter we explicitly integrate each term in the binomial sum leading to
\begin{equation}
-\sum_{j=1}^{n_t-1}\left(\begin{array}{c} n_t-1 \\ j \end{array}\right)\frac{(-1)^j}{j}=\gamma+\Psi(n_t).
\end{equation}
Thus, the sum rate becomes
\begin{equation}
\begin{split}
R=\frac{n_t}{\delta}\int_{1}^\infty dx({\rm ln}(x)+\gamma+\Psi(n_t)) K\left(-\frac{1}{\delta},\frac{x-1}{\delta}\right).
\end{split}
\end{equation}
The integration variable $x$ can be substituted to $x\to \delta x+1$ and we may exploit the explicit representation~\eqref{kernel} and the orthogonality relation  to perform the integration over the constant $\gamma+\Psi(n_t)$. This leads to our second statement~\eqref{sumrateMMSE}.
\end{IEEEproof}

Again we want to give some examples of these results. For this purpose we consider the P\'olya ensembles anew, see~\eqref{weightp}-\eqref{kernelPolya}.  For the distribution of the SINR, the two integrals over  $u$ in~\eqref{kernelPolya} and $x$ in~\eqref{distributionMMSE} can be interchanged. Then, we need to perform the following integral
\begin{equation}
\begin{split}
g=&\int_0^1 dx (1-x)^{n_t-2}\left(x+\frac{1}{\gamma^{\rm mmse}}\right)(-\partial_x)^{n_t}[x^{n_t}\omega(x \hat{u})]\\
=&(-1)^{n_t}(n_t-2)!\left[\left(\frac{n}{\gamma^{\rm mmse}}+1\right)\omega(\hat{u})+\hat{u}\partial_{\hat{u}}\omega(\hat{u})\right],
\end{split}
\end{equation}
which is integrated by parts. We set now $\hat{u}=\gamma^{\rm mmse} u/\delta$, and the distribution of the SINR translates to
\begin{equation}
\begin{split}
&\rho_{\rm P\text{\'o}lya}(\gamma^{\rm mmse})=-\left(\frac{-\gamma^{\rm mmse}}{\gamma^{\rm mmse}+1}\right)^{n_t}\int_{0}^1 \frac{du\,p_{n_t-1}\left(-u/\delta\right)}{\delta\,\mathcal{M}\omega(n_t)}\\
&\times\left[\left(\frac{n_t}{\gamma^{\rm mmse}}+1\right) \omega\left(\frac{\gamma^{\rm mmse}}{\delta}u\right)+u\partial_u\omega\left(\frac{\gamma^{\rm mmse}}{\delta}u\right)\right].
\end{split}
\end{equation}
For the sum rate there is not such a simple formula. Nevertheless, we can  find simple sums for a channel matrix $\Hm=\Hm_M\cdots \Hm_1$ that is a product for normal distributed rectangular matrices. For the distribution of the SINR, the expression is
\begin{equation}\label{ProddistributionMMSE}
\begin{split}
\rho_{M}(\gamma^{\rm mmse})
=&\left(\frac{\gamma^{\rm mmse}}{\gamma^{\rm mmse}+1}\right)^{n_t-1}\sum_{j=0}^{n_t-1}\left(\begin{array}{c} n_t-1 \\ j \end{array}\right)\frac{\delta^{-j-1}}{\prod_{l=1}^M(\nu_l+j)!}\\
\times& G_{1,M+1}^{M,1}\left(\left.\begin{array}{c} -j,-\frac{\gamma^{\rm mmse}+n_t}{\gamma^{\rm mmse}+1};\, - \\ \nu_1,\ldots,\nu_M;-1-j,1-\frac{\gamma^{\rm mmse}+n_t}{\gamma^{\rm mmse}+1} \end{array}\right|\frac{\gamma^{\rm mmse}}{\delta}\right).
\end{split}
\end{equation}
For the sum rate we can apply the convolution of two Meijer G-function when exploiting the identity~\eqref{log}, so that after interchanging the $u$ and the $x$ integral in~\eqref{sumrateMMSE} and~\eqref{kernelPolya}, we have
\begin{equation}\label{sumrateMMSEGin}
\begin{split}
&R_{M}=n_t(\gamma+\Psi(n_t))+\sum_{j=0}^{n_t-1}\frac{n_t\delta^{-j-1}}{j!(n_t-1-j)!\prod_{l=1}^M(\nu_l+j)!}\\
&\times G_{4,M+4}^{M+2,3}\left(\left.\begin{array}{c} -1,-n_t,-j;0 \\ -1,-1,\nu_1,\ldots,\nu_M;0,-1-j\end{array}\right|\frac{1}{\delta}\right).
\end{split}
\end{equation}
We can find similar results when choosing a product of truncated unitary matrices (e.g. a model exploited for optical communications) instead of normal distributed matrices.

\section{Numerical Comparisons}\label{sec:numerics}

In this section, analytical findings are contrasted with Monte Carlo simulations; the reference scenario is a MIMO channel affected by progressive scattering through $M=3$ independent scattering clusters, with uncorrelated antennas at both transmit and receive side.

\begin{figure}[t!]
\centerline{\includegraphics[width=0.8\textwidth]{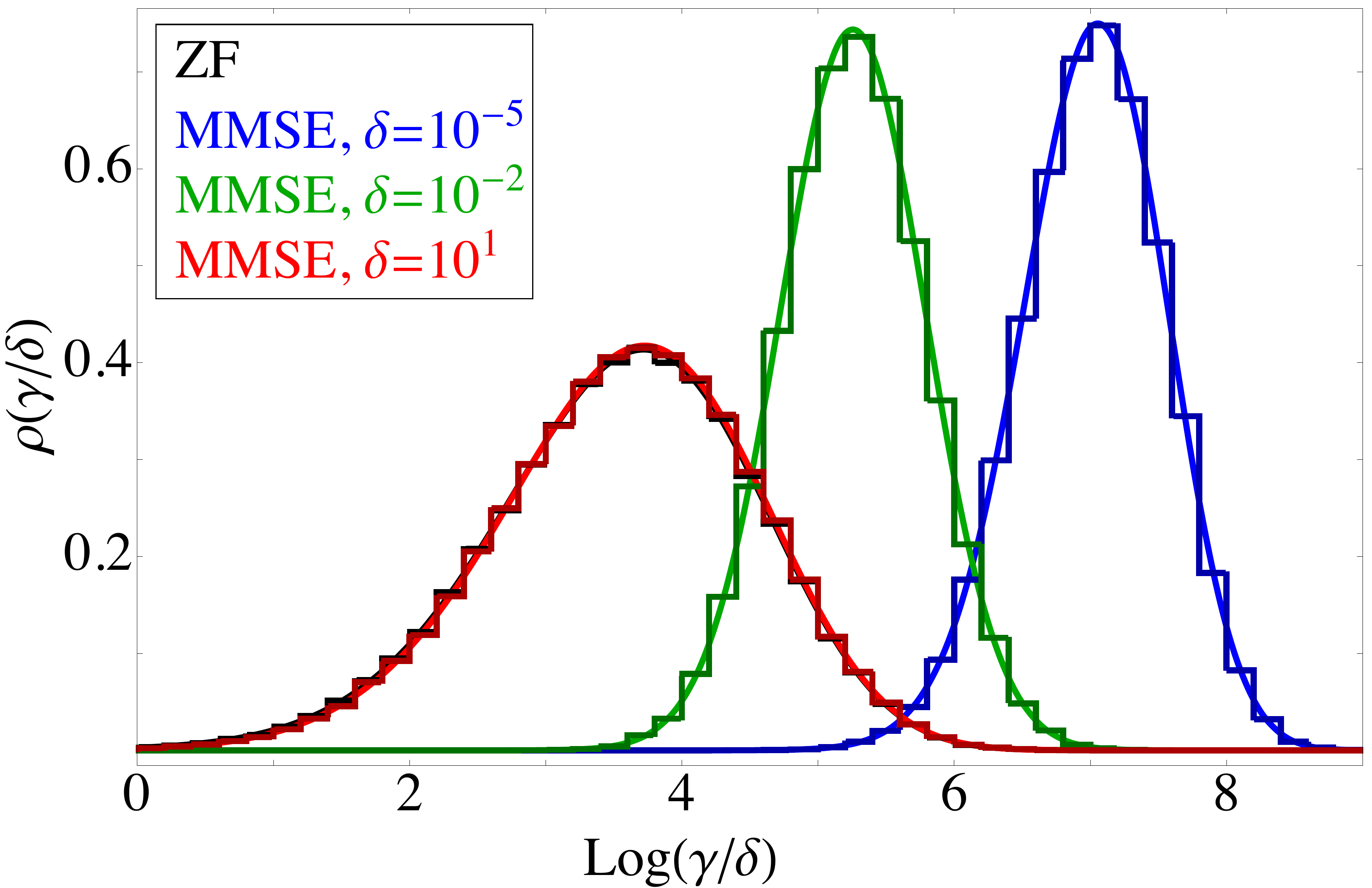}
}
\caption{Pdf of the output SINR on a progressive scattering channel with $n_t=8$, $M=3$, and $(\nu_1,\nu_2,\nu_3)=(5,2,2)$,  for ZF receivers (black) and MMSE receivers, with $\delta$ as a parameter. The analytical expressions~\eqref{ProddistributionZF} and~\eqref{ProddistributionMMSE} (solid curves) are compared to Monte Carlo simulations (histograms) based on $10^5$ samples. For $\delta=10$, ZF and MMSE densities almost perfectly overlap. Note that
 due to the rescaling of the SINR $\gamma$ by $\delta$ the dependence on $\delta$ is lost for ZF receivers, cf. \eqref{ithmode}.}
\label{fig:density}
\end{figure}

\begin{figure}[t!]
\centerline{\includegraphics[width=0.8\textwidth]{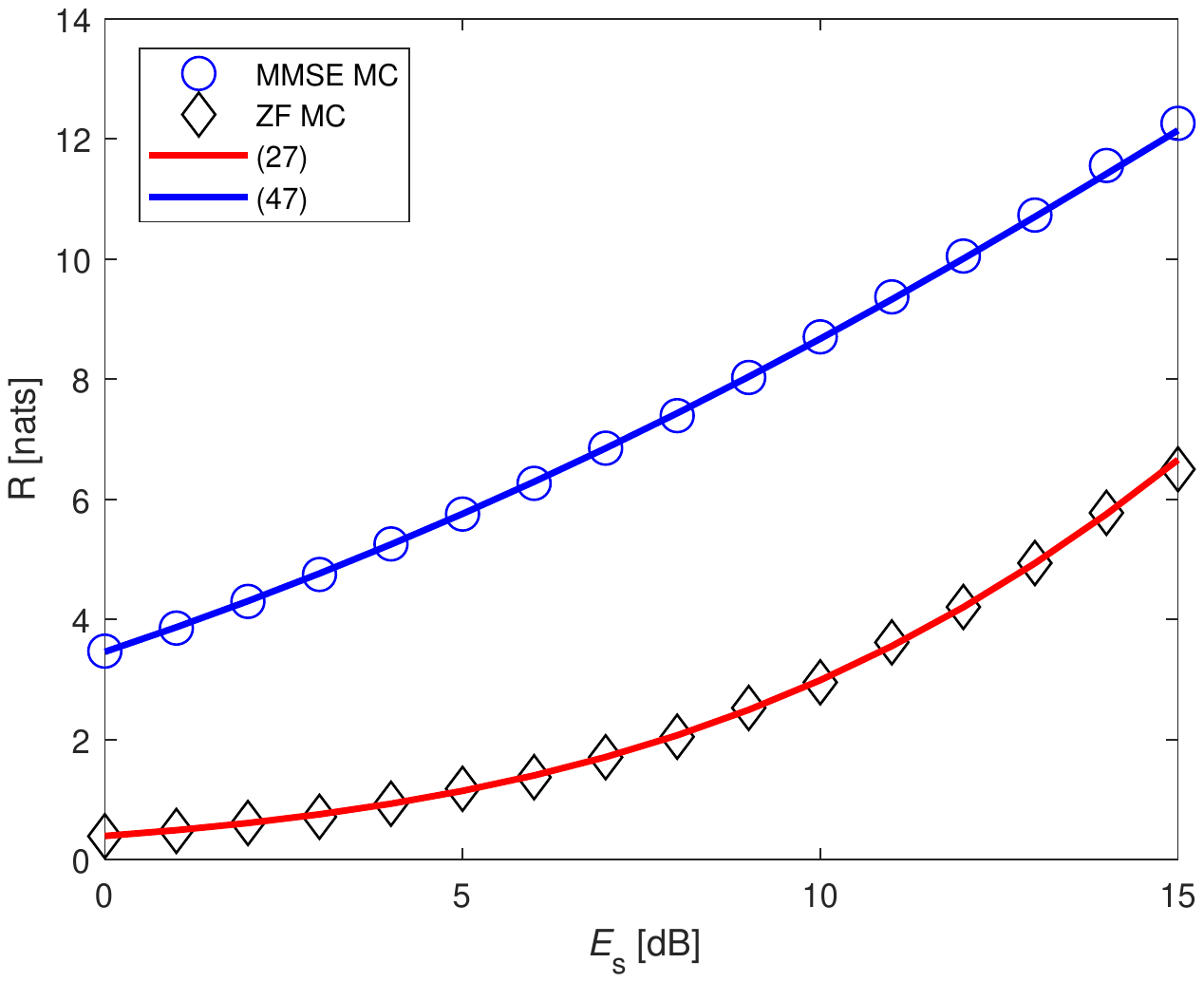}
}
\caption{The sum rate $R$ for ZF receivers (red curve) and MMSE receivers (blue curve), as a function of the input power $\Ec_s$ in dB.  The random matrix ensemble is the same as in figure~\ref{fig:density}. The channel is the same as in Figure~\ref{fig:density}. The symbols represent the Monte Carlo simulations while the solid curves are the analytical results~\eqref{ProdsumrateZF} and~\eqref{sumrateMMSEGin}.We recall
 that $\delta=\mathcal{E}_s \alpha/n_t$ where $\alpha/n_t\approx 10^{-3}$ in our setting.}
\label{fig:sumrate}
\end{figure}

Figure~\ref{fig:density} shows the output SINR densities for both the analyzed linear receivers,  for increasing values of  $\delta$.
Notice that because of plotting the pdfs as a function of $\gamma/\delta$ the pdf of ZF is $\delta$-independent.
For large $\delta=10$ the curves for MMSE and ZF become indistinguishable, as can be expected from \eqref{ithmode}.
In Figure~\ref{fig:sumrate}, the ergodic sum rate under the same fading assumptions is plotted, as a function of the  input SNR $\Ec_s$. Recall that, as defined in Section\ref{sec:model}, $\delta=\frac{\Ec_s\alpha}{n_t}$. This corresponds to a uniform power allocation among transmit antennas (i.e. each antenna is fed by $\Ec_s/n_t$).  Markers represent Monte Carlo simulations (obtained with $1000$ runs each) and continuous curves correspond to~\eqref{ProdsumrateZF} (blue)  and~\eqref{sumrateMMSEGin} (dark), respectively. On the considered input power range, MMSE outperforms ZF.

\section{Conclusions}\label{sec:conclusio}

Probability density functions of the output SINR of ZF and MMSE MIMO receivers are derived for a broad class of fading scenarios, encompassing both wireless as well as optical transmission links. As a first application of our newly derived results, the expected value of the sum rate achievable with independent stream decoding is expressed in closed form. Depending on the assumed fading law, more compact or previously unavailable expressions are provided. The link with the correlation kernel of the eigenvalues density of the channel matrix made explicit will allow a simplified analysis of massive MIMO settings with ZF precoding/reception. Further applications of our results can be devised in the energy-efficient design of multiuser wireless systems and in the comparative performance analysis of linear filters. In particular, our findings can help in  extending the class of fading laws for which, e.g., the SINR gap between MMSE and ZF can be explicitly analyzed, and in analyzing separately  the impact of different sets of spatial degrees of freedom (e.g. number of scattering clusters versus number of transmit/receive antennas) on the overall performance.

\end{document}